\documentclass[aps,prd,reprint,showpacs,twoside]{revtex4-1}

\usepackage{mathrsfs}
\usepackage{bm}
\usepackage{amsfonts}
\usepackage{amsmath}
\usepackage{amssymb}
\usepackage{graphicx}
\usepackage[colorlinks=true,linkcolor=blue,citecolor=blue,breaklinks,linktoc=page,urlcolor=blue]{hyperref}

\begin{document}
\title{Bimetric Gravity From Adjoint Frame Field In Four Dimensions}

\author{Zhi-Qiang Guo}\email{zhiqiang.guo@usm.cl}\affiliation{Departamento de F\'{i}sica y Centro Cient\'{i}fico
Tecnol\'{o}gico de Valpara\'{i}so,\\ Universidad T\'{e}cnica Federico
Santa Mar\'{i}a,\\ Casilla 110-V, Valpara\'{i}so, Chile}
\author{Iv\'{a}n Schmidt}\email{ivan.schmidt@usm.cl}\affiliation{Departamento de F\'{i}sica y Centro Cient\'{i}fico
Tecnol\'{o}gico de Valpara\'{i}so,\\ Universidad T\'{e}cnica Federico
Santa Mar\'{i}a,\\ Casilla 110-V, Valpara\'{i}so, Chile}

\begin{abstract}
We provide a novel model of gravity by using adjoint frame fields in four dimensions. It has a natural interpretation as a gravitational theory of a complex metric field, which describes interactions between two real metrics. The classical solutions establish three appealing features. The spherical symmetric black hole solution has an additional hair, which includes the Schwarzschild solution as a special case. The de Sitter solution is realized without introducing a cosmological constant. The constant flat background breaks the Lorentz invariance spontaneously, although the Lorentz breaking effect can be localized to the second metric while the first metric still respects the Lorentz invariance.
\end{abstract}


\maketitle

\section{Introduction}\label{sec1}

Since its construction in 1915, Einstein's gravitational theory has passed numerous experimental tests in the last century. However, the observation of cosmological acceleration at later times provided us with the hint that Einstein's gravity could need to be modified at large distances, and in fact several modifications have been proposed  along different directions~\cite{Brans:1961sx,Sotiriou:2008rp,Dvali:2000hr,Nicolis:2008in,Horndeski:1974wa,Deffayet:2009mn,Horava:2009uw,Milgrom:2009gv}. One intriguing possibility is that there could exist another metric field ~$f_{\mu\nu}$ besides the conventional metric field~$g_{\mu\nu}$. This kind of bimetric gravity was proposed by Rosen in 1940~\cite{Rosen:1940zza}, although in general it has been shown to present a ghost problem~\cite{Boulware:1973my}.  Nevertheless, the ghost-free bimetric gravity has recently been constructed~\cite{deRham:2010kj,Hassan:2011zd} in the framework of massive gravity~\cite{deRham:2014zqa}, which is viable to explain the cosmological acceleration~\cite{Akrami:2015qga}.

In this paper we propose a novel model of bimetric gravity, which is not within those traditionally considered. In Einstein's theory, gravity is described by the metric tensor~$g_{\mu\nu}$, which can be recast into an equivalent formulation by using the frame field $e^{I}_{\mu}$. Here $e^{I}_{\mu}$ transforms as the fundamental representation of the local Lorentz group $SO(1,3)$. Alternatively, one may wonder if consistent gravitational theories exist based on the adjoint representation $e^{IJ}_{\mu}$. An immediate consequence of this adjoint frame field is that we can obtain two gauge invariant metric fields~\footnote{Here we use the convention that $\eta=\mathrm{diag}(-1,1,1,1)$, and~$\epsilon^{0123}=1$.}
\begin{eqnarray}
\label{sec1-g}
g_{\mu\nu}=\frac{1}{2}\eta_{IM}\eta_{JN}e^{IJ}_{\mu}e^{MN}_{\nu}
\end{eqnarray}
and
\begin{eqnarray}
\label{sec1-f}
f_{\mu\nu}=\frac{1}{4}\epsilon_{IJMN}e^{IJ}_{\mu}e^{MN}_{\nu}.
\end{eqnarray}
Hence a theory of  $e^{IJ}_{\mu}$ intrinsically describes interactions between two metrics. In contrast with the massive bimetric gravity, which retains intact the Lorentz invariance~\cite{deRham:2010kj,Hassan:2011zd}, the bimetric theory of  $e^{IJ}_{\mu}$ has to break Lorentz covariance spontaneously. One way to see this is by counting degrees of freedom~(dofs):  $e^{IJ}_{\mu}$ has $18$ gauge invariant dofs, but $g_{\mu\nu}$ and $f_{\mu\nu}$ have a total of $20$ superficial dofs. So $g_{\mu\nu}$ and $f_{\mu\nu}$ are not independent, and actually they are constrained by the complex condition
\begin{eqnarray}
\label{sec1-gf}
\mathrm{det}(g_{\mu\nu}+i{f}_{\mu\nu})=0,
\end{eqnarray}
which eliminates two dofs.  This constraint also forbids $g_{\mu\nu}$ and $f_{\mu\nu}$ to be proportional to the Lorentz metric simultaneously, hence at least one of them is required to break Lorentz invariance. Nevertheless, as demonstrated in~\cite{Rubakov:2008nh,Dubovsky:2004sg,Gabadadze:2004iv,Blas:2007ep,Comelli:2013txa}, the spontaneous breaking of Lorentz invariance is not problematic, but it is useful in order to establish the consistence of bimetric gravity.

In section~\ref{sec2}, we construct the first order Lagrangian, using the frame fields. We show that a concise formulation of the Lagrangian exists when we use the $SO(3,\mathbb{C})$ variables instead of the $SO(1,3)$ variables, and that the spin connection can be uniquely determined thorough the variation principle. In section~\ref{sec3}, we provide the metric-like formulation, using the complex metric fields $g_{\mu\nu}+i{f}_{\mu\nu}$. In section~\ref{sec4}, we obtain the spherical black hole solution and the time-dependent solution.

\section{Frame-like Formulation}\label{sec2}

We begin with the Frame-like formulation. Consider the frame field $e^{IJ}_{\mu}$, where $I, J=0,1,2,3$ are the indices of $SO(1,3)$ adjoint representation, and $e^{IJ}_{\mu}$ is antisymmetric with respect to $I$ and $J$. The field strength is built from the spin connection
\begin{eqnarray}
\label{sec2-fs}
F^{IJ}_{\mu\nu}=\partial_{\mu}\omega^{IJ}_{\nu}-\partial_{\nu}\omega^{IJ}_{\mu}
+\omega^{I}_{\mu{N}}\omega^{NJ}_{\nu}-\omega^{I}_{\nu{N}}\omega^{NJ}_{\mu}.
\end{eqnarray}
Besides $g_{\mu\nu}$ and $f_{\mu\nu}$ in Eqs.~(\ref{sec1-g}) and~(\ref{sec1-f}), we also have two other gauge invariant quantities
\begin{eqnarray}
\label{sec2-u}
u^{\mu}&=&\frac{1}{6}\epsilon^{\mu\nu\rho\sigma}\eta_{IK}\eta_{JL}\eta_{MN}e^{IM}_{\nu}e^{NJ}_{\rho}e^{KL}_{\sigma},\\
\label{sec2-v}
v^{\mu}&=&-\frac{1}{12}\epsilon^{\mu\nu\rho\sigma}\epsilon_{IJMN}\eta_{KL}e^{IJ}_{\nu}e^{MK}_{\rho}e^{NL}_{\sigma}.
\end{eqnarray}
From $e^{IJ}_{\mu}$ and $F^{IJ}_{\mu\nu}$, we can obtain the following gauge invariant operators
\begin{eqnarray}
\label{sec2-s-1}
\mathcal{O}_{1}&=&{^{\ast}F^{\mu\nu}_{IJ}}e^{IM}_{\mu}e^{NJ}_{\nu}\eta_{MN},\\
\label{sec2-s-2}
\mathcal{O}_{2}&=&{^{\ast}F^{\mu\nu{IJ}}}e^{KM}_{\mu}e^{NL}_{\nu}\epsilon_{IJKL}\eta_{MN}.
\end{eqnarray}

We also have the operators associated to $u^{\alpha}$
\begin{eqnarray}
\label{sec2-s-3}
O_{1}&=&{^{\ast}F^{\mu\nu}_{IJ}}e^{IJ}_{\nu}u^{\rho}g_{\rho\mu},
O_{2}={^{\ast}F^{\mu\nu}_{IJ}}e_{\nu{KL}}\epsilon^{IJKL}u^{\rho}g_{\rho\mu},\\
O_{3}&=&{^{\ast}F^{\mu\nu}_{IJ}}e^{IJ}_{\nu}u^{\rho}f_{\rho\mu},
O_{4}={^{\ast}F^{\mu\nu}_{IJ}}e_{\nu{KL}}\epsilon^{IJKL}u^{\rho}f_{\rho\mu},\nonumber
\end{eqnarray}
and the operators associated to $v^{\alpha}$
\begin{eqnarray}
\label{sec2-s-4}
O_{5}&=&{^{\ast}F^{\mu\nu}_{IJ}}e^{IJ}_{\nu}v^{\rho}g_{\rho\mu},
O_{6}={^{\ast}F^{\mu\nu}_{IJ}}e_{\nu{KL}}\epsilon^{IJKL}v^{\rho}g_{\rho\mu},~\\
O_{7}&=&{^{\ast}F^{\mu\nu}_{IJ}}e^{IJ}_{\nu}v^{\rho}f_{\rho\mu},
O_{8}={^{\ast}F^{\mu\nu}_{IJ}}e_{\nu{KL}}\epsilon^{IJKL}v^{\rho}f_{\rho\mu},\nonumber
\end{eqnarray}
where ${^{\ast}F^{\mu\nu}_{IJ}}=\epsilon^{\mu\nu\alpha\beta}F_{\alpha\beta{IJ}}$ is the dual field strength. In order to construct the Lagrangian, we also need the gauge invariant pseudo scalars and scalars
\begin{eqnarray}
\label{sec2-s-5}
g&=&\mathrm{det}(g_{\mu\nu}),~~f=\mathrm{det}(f_{\mu\nu}),\\
\label{sec2-s-6}
\varrho&=&f^{\mu\nu}g_{\mu\nu},~~\varsigma=g^{\mu\nu}f_{\mu\nu}.
\end{eqnarray}

As defined in Eqs.~(\ref{sec1-g}) and~(\ref{sec1-f}), $g_{\mu\nu}$ and $f_{\mu\nu}$ are in general invertible, and $g^{\mu\nu}$ and $f^{\mu\nu}$ are their inverses respectively. Then a general Lagrangian can be constructed as
\begin{eqnarray}
\label{sec2-lag-re}
\mathcal{L}=a_{1}\mathcal{O}_{1}+a_{2}\mathcal{O}_{2}+\sum_{i=1}^{8}o_{i}\chi_{i}O_{i},
\end{eqnarray}
where $a_{i}$ and $o_{i}$ are constant coefficients, and $\chi_{i}(g,f,\varrho,\varsigma)$ are some pseudo-scalar functions, to ensure that $\mathcal{L}$ has the proper transformation properties.  The Lagrangian~(\ref{sec2-lag-re}) is constructed using the $SO(1,3)$ variables $e^{IJ}_{\mu}$ and $\omega^{IJ}_{\mu}$, and a general Lagrangian can depend on $10$ constant coefficients and $8$ pseudoscalar functions. However, in the following we shall show that a simplified version can be constructed if we use the $SO(3,\mathbb{C})$ variables. We define the complex variables
\begin{eqnarray}
\label{sec2-e-c}
\mathbf{e}^{k}_{\mu}&=&-\frac{1}{2}\epsilon^{kmn}{e}^{mn}_{\mu}+i{e}^{0k}_{\mu},\\
\label{sec2-w-c}
\mathbf{A}^{k}_{\mu}&=&-\frac{1}{2}\epsilon^{kmn}{\omega}^{mn}_{\mu}+i{\omega}^{0k}_{\mu},
\end{eqnarray}
where the small latin letters take values  $1,2,3$. Then $\mathbf{e}^{k}_{\mu}$ and $\mathbf{A}^{k}_{\mu}$ are the $SO(3,\mathbb{C})$ variables. We can also define the complex field strength
\begin{eqnarray}
\label{sec2-fs-c}
\mathbf{F}^{k}_{\mu\nu}&=&-\frac{1}{2}\epsilon^{kmn}{F}^{mn}_{\mu\nu}+i{F}^{0k}_{\mu\nu},
\end{eqnarray}
which can be rewritten in terms of the complex connection as
\begin{eqnarray}
\label{sec2-fs-c-def}
\mathbf{F}^{k}_{\mu\nu}&=&\partial_{\mu}\mathbf{A}^{k}_{\nu}-\partial_{\nu}\mathbf{A}^{k}_{\mu}
+\epsilon^{kmn}\mathbf{A}^{m}_{\mu}\mathbf{A}^{n}_{\nu}.
\end{eqnarray}

Using the complex variables, we have the complex metric
\begin{eqnarray}
\label{sec2-g-c}
\mathbf{g}_{\mu\nu}&=&\mathbf{e}^{k}_{\mu}\mathbf{e}^{k}_{\nu}=g_{\mu\nu}+i{f}_{\mu\nu}.
\end{eqnarray}
and the complex pseudo-vector
\begin{eqnarray}
\label{sec2-u-c}
\mathbf{u}^{\mu}&=&\frac{1}{6}\epsilon^{\mu\nu\rho\sigma}\epsilon^{kmn}\mathbf{e}^{k}_{\nu}\mathbf{e}^{m}_{\rho}\mathbf{e}^{n}_{\sigma}
=u^{\mu}+i{v}^{\mu}.
\end{eqnarray}
From Eqs.~(\ref{sec2-g-c}) and~(\ref{sec2-u-c}), we see the two facets of $\mathbf{g}_{\mu\nu}$ and $\mathbf{u}^{\mu}$. They can be expressed either using the $SO(3,\mathbb{C})$ variable in Eq.~(\ref{sec2-e-c}) or using the $SO(1,3)$ variables in Eqs.~(\ref{sec1-g}), (\ref{sec1-f}), (\ref{sec2-u}) and (\ref{sec2-v}). From Eq.~(\ref{sec2-g-c}), we know that the determinant of $\mathbf{g}_{\mu\nu}$ is zero, and we can also verify that $\mathbf{u}^{\mu}$ is its eigenvector with eigenvalue $0$. Actually, $\mathbf{g}_{\mu\nu}$ and $\mathbf{u}^{\mu}$ satisfy the identities
\begin{eqnarray}
\label{sec2-u-g-c-1}
\mathbf{g}_{\mu\rho}\mathbf{u}^{\rho}&=&0,\\
\label{sec2-u-g-c-2}
\mathbf{u}^{\mu}\mathbf{u}^{\nu}
&=&\frac{1}{6}\epsilon^{\mu\alpha\rho\tau}\epsilon^{\nu\beta\sigma\theta}
\mathbf{g}_{\alpha\beta}\mathbf{g}_{\rho\sigma}\mathbf{g}_{\tau\theta}.
\end{eqnarray}
For a given $\mathbf{g}_{\mu\nu}$, we can obtain $\mathbf{u}^{\mu}$ by using Eq.~(\ref{sec2-u-g-c-2}). So $\mathbf{u}^{\mu}$ is not independent of $\mathbf{g}_{\mu\nu}$, but it can be completely determined by $\mathbf{g}_{\mu\nu}$. We also define a complex pseudo-scalar
\begin{eqnarray}
\label{sec2-u-g-sca}
\Phi=\frac{1}{4}\mathbf{\bar{g}}_{\mu\nu}\mathbf{u}^{\mu}\mathbf{u}^{\nu}=g-f+i{g}\varsigma,
\end{eqnarray}
where $\mathbf{\bar{g}}_{\mu\nu}$ is the complex conjugate of the complex metric $\mathbf{g}_{\mu\nu}$. Using $\Phi$, we define the covariant pseudo-vector
\begin{eqnarray}
\label{sec2-v-g}
\mathbf{v}_{\mu}=-\frac{1}{4\Phi}\mathbf{\bar{g}}_{\mu\rho}\mathbf{u}^{\rho},
\end{eqnarray}
which satisfies
\begin{eqnarray}
\label{sec2-v-u-g}
\mathbf{u}^{\alpha}\mathbf{v}_{\alpha}=-1
\end{eqnarray}
according to Eq.~(\ref{sec2-u-g-sca}). Using $\mathbf{v}_{\mu}$, we can define
\begin{eqnarray}
\label{sec2-e-inv}
\mathbf{E}^{\alpha}_{k}=\frac{1}{2}\epsilon^{\alpha\theta\rho\sigma}\epsilon^{kmn}
\mathbf{v}_{\theta}\mathbf{e}^{m}_{\rho}\mathbf{e}^{n}_{\sigma},
\end{eqnarray}
then we have
\begin{eqnarray}
\label{sec2-e-inv-e}
\mathbf{E}^{\alpha}_{j}\mathbf{e}^{i}_{\alpha}=\delta^{i}_{j},
\end{eqnarray}
and we also have
\begin{eqnarray}
\label{sec2-v-inv-e-n}
\mathbf{n}^{\alpha}_{\beta}=\mathbf{E}^{\alpha}_{k}\mathbf{e}^{k}_{\beta}.
\end{eqnarray}
$\mathbf{E}^{\alpha}_{i}$ can be regarded as the left inverse of $\mathbf{e}^{i}_{\alpha}$, and $\mathbf{n}^{\alpha}_{\beta}$ plays the role of the projection tensor. Note that $\mathbf{g}_{\mu\nu}$ is not invertible. We can obtain its generalized inverse as
\begin{eqnarray}
\label{sec2-v-g-inv}
\mathbf{g}^{\mu\nu}=\mathbf{E}^{\mu}_{k}\mathbf{E}^{\nu}_{k}=\frac{1}{2}\epsilon^{\mu\alpha\rho\tau}\epsilon^{\nu\beta\sigma\theta}
\mathbf{v}_{\alpha}\mathbf{v}_{\beta}\mathbf{g}_{\rho\sigma}\mathbf{g}_{\tau\theta},
\end{eqnarray}
which satisfies
\begin{eqnarray}
\label{sec2-v-g-inv-re-1}
\mathbf{g}_{\mu\rho}\mathbf{g}^{\rho\sigma}\mathbf{g}_{\sigma\nu}&=&\mathbf{g}_{\mu\nu},
\mathbf{g}^{\mu\rho}\mathbf{g}_{\rho\sigma}\mathbf{g}^{\sigma\nu}=\mathbf{g}^{\mu\nu}.
\end{eqnarray}
And we also have the relation
\begin{eqnarray}
\label{sec2-v-g-inv-re-2}
\mathbf{n}^{\alpha}_{\beta}=\mathbf{g}^{\alpha\rho}\mathbf{g}_{\rho\beta}
=\delta^{\alpha}_{\beta}+\mathbf{u}^{\alpha}\mathbf{v}_{\beta}.
\end{eqnarray}
 Using the above complex variables, we propose the Lagrangian
\begin{eqnarray}
\label{sec2-lag-c}
\mathscr{L}&=&
\frac{\kappa}{4}\epsilon^{\mu\nu\alpha\beta}\epsilon^{ijk}\mathbf{e}^{i}_{\mu}\mathbf{e}^{j}_{\nu}\mathbf{F}^{k}_{\alpha\beta}\\
&+&\lambda\sqrt{-\Phi}\epsilon^{\mu\nu\alpha\beta}\mathbf{v}_{\mu}\mathbf{e}^{k}_{\nu}\mathbf{F}^{k}_{\alpha\beta}
+\mathrm{c.c},\nonumber
\end{eqnarray}
where $\kappa$ and $\lambda$ are complex constants, and $\mathrm{c.c}$ is the complex conjugate. Compared with Eq.~(\ref{sec2-lag-re}), Eq.~(\ref{sec2-lag-c}) has $4$ real constant coefficients. It is a special case of Eq.~(\ref{sec2-lag-re}), and it can be rewritten into the formulation~(\ref{sec2-lag-re}) in terms of the $SO(1,3)$ variables. In the following, we shall consider the concise formulation~(\ref{sec2-lag-c}) in terms of $SO(3,\mathbb{C})$ variables.  Varying  Eq.~(\ref{sec2-lag-c}) with respect to $\mathbf{A}^{i}_{\mu}$, we obtain
\begin{eqnarray}
\label{sec2-lag-tor-eq}
\kappa\epsilon^{\mu\nu\alpha\beta}\epsilon^{ijk}\mathbf{T}^{j}_{\mu\alpha}\mathbf{e}^{k}_{\beta}
&=&2\lambda\sqrt{-\Phi}\epsilon^{\nu\mu\alpha\beta}\mathbf{v}_{\alpha}\mathbf{T}^{i}_{\mu\beta}\nonumber\\
&+&2\lambda\epsilon^{\nu\mu\alpha\beta}\partial_{\mu}(\mathbf{v}_{\alpha}\sqrt{-\Phi})\mathbf{e}^{i}_{\beta},
\end{eqnarray}
where
\begin{eqnarray}
\label{sec2-lag-tor}
\mathbf{T}^{i}_{\mu\nu}
=\frac{1}{2}(\partial_{\mu}\mathbf{e}^{i}_{\nu}+\epsilon^{ijk}\mathbf{A}^{j}_{\mu}\mathbf{e}^{k}_{\nu})
-(\mu\leftrightarrow\nu)
\end{eqnarray}
is the torsion tensor. Variations of Eq.~(\ref{sec2-lag-c}) with respect to $\mathbf{e}^{i}_{\alpha}$ yield
\begin{eqnarray}
\label{sec2-lag-eq}
&-&\frac{\kappa}{2}\epsilon^{\mu\nu\alpha\theta}\epsilon^{ijk}\mathbf{e}^{j}_{\theta}\mathbf{F}^{k}_{\mu\nu}
+\lambda\sqrt{-\Phi}\epsilon^{\mu\nu\alpha\beta}\mathbf{v}_{\beta}\mathbf{F}^{i}_{\mu\nu}\nonumber\\
&+&\frac{\lambda}{4\sqrt{-\Phi}}\epsilon^{\mu\nu\rho\sigma}\mathbf{e}^{k}_{\sigma}\mathbf{F}^{k}_{\mu\nu}
\mathbf{\bar{g}}_{\rho\theta}
\mathbf{g}^{\theta\tau}\mathbf{u}^{\alpha}\mathbf{e}^{i}_{\tau}\\
&=&
\frac{\bar{\lambda}}{4\sqrt{-\overline{\Phi}}}\epsilon^{\mu\nu\rho\sigma}
\mathbf{\bar{e}}^{k}_{\sigma}\mathbf{\bar{F}}^{k}_{\mu\nu}
\bigg(\mathbf{\bar{u}}^{\alpha}\mathbf{\bar{u}}^{\theta}\mathbf{\bar{v}}_{\rho}
+\mathbf{\bar{u}}^{\alpha}\delta^{\theta}_{\rho}+\mathbf{\bar{u}}^{\theta}\delta^{\alpha}_{\rho}\bigg)\mathbf{e}^{i}_{\theta}.\nonumber
\end{eqnarray}
In the above, $\bar{z}$ always means the complex conjugate of the complex variable $z$. Eqs.~(\ref{sec2-lag-tor-eq}) and~(\ref{sec2-lag-eq}) give the first order formulation of the equations of motion.

\section{Metric-like Formulation}\label{sec3}

In the above, we have obtained the first order formulation. In this section, we show that the second order formulation gives simpler expressions in terms of metric variables. First we need to solve Eq.~(\ref{sec2-lag-tor-eq}) to obtain the connection. We define the affine connection
\begin{eqnarray}
\label{sec3-lag-con}
\mathbf{\Gamma}^{i}_{\mu\nu}=\partial_{\mu}\mathbf{e}^{i}_{\nu}+\epsilon^{ijk}\mathbf{A}^{j}_{\mu}\mathbf{e}^{k}_{\nu}.
\end{eqnarray}
The metric-like formulation of this connection is
\begin{eqnarray}
\label{sec3-lag-con-met}
\mathbf{\Gamma}^{\rho}_{\mu\nu}=\mathbf{E}_{i}^{\rho}\mathbf{\Gamma}^{i}_{\mu\nu},
\end{eqnarray}
which satisfy the condition
\begin{eqnarray}
\label{sec3-lag-con-met-s}
\mathbf{\Gamma}^{\rho}_{\mu\nu}=\mathbf{n}^{\rho}_{\tau}\mathbf{\Gamma}^{\tau}_{\mu\nu}
\end{eqnarray}
according to Eqs.~(\ref{sec2-e-inv-e}) and (\ref{sec2-v-inv-e-n}). From Eq.~(\ref{sec3-lag-con-met}), we have
\begin{eqnarray}
\label{sec3-lag-con-re}
\partial_{\mu}\mathbf{e}^{i}_{\nu}+\epsilon^{ijk}\mathbf{A}^{j}_{\mu}\mathbf{e}^{k}_{\nu}
=\mathbf{\Gamma}^{\rho}_{\mu\nu}\mathbf{e}^{i}_{\rho}.
\end{eqnarray}
From this equation, we obtain the metric compatibility condition
\begin{eqnarray}
\label{sec3-lag-g}
\partial_{\mu}\mathbf{g}_{\alpha\beta}
=\mathbf{\Gamma}^{\rho}_{\mu\alpha}\mathbf{g}_{\rho\beta}+\mathbf{\Gamma}^{\rho}_{\mu\beta}\mathbf{g}_{\rho\alpha}.
\end{eqnarray}
This equation can be solved as
\begin{eqnarray}
\label{sec3-lag-con-sol}
\mathbf{\Gamma}^{\rho}_{\mu\nu}=\mathbf{g}^{\rho\sigma}\mathbf{\Gamma}_{\sigma,\mu\nu}
-\mathbf{g}^{\rho\sigma}(\mathbf{T}^{\theta}_{\mu\sigma}\mathbf{g}_{\theta\nu}+\mathbf{T}^{\theta}_{\nu\sigma}\mathbf{g}_{\theta\mu})
+\mathbf{T}^{\rho}_{\mu\nu},~~
\end{eqnarray}
where
\begin{eqnarray}
\label{sec3-lag-tor}
\mathbf{T}^{\rho}_{\mu\nu}=\frac{1}{2}(\mathbf{\Gamma}^{\rho}_{\mu\nu}-\mathbf{\Gamma}^{\rho}_{\nu\mu}),
\end{eqnarray}
is the torsion tensor, and
\begin{eqnarray}
\label{sec3-lag-tor-ch}
\mathbf{\Gamma}_{\tau,\rho\sigma}=\frac{1}{2}(\partial_{\rho}\mathbf{g}_{\tau\sigma}+\partial_{\sigma}\mathbf{g}_{\tau\rho}
-\partial_{\tau}\mathbf{g}_{\rho\sigma})
\end{eqnarray}
is the Christoffel symbol of the first kind. In order to satisfy Eq.~(\ref{sec3-lag-g}), the torsion tensor need to satisfy
\begin{eqnarray}
\label{sec3-lag-tor-s}
\mathbf{u}^{\tau}\mathbf{\Gamma}_{\tau,\alpha\beta}=
\mathbf{u}^{\tau}(\mathbf{T}^{\rho}_{\alpha\tau}\mathbf{g}_{\rho\beta}+\mathbf{T}^{\rho}_{\beta\tau}\mathbf{g}_{\rho\alpha}).
\end{eqnarray}
Using the torsion tensor, Eq.~(\ref{sec2-lag-tor-eq}) can be rewritten as
\begin{eqnarray}
\label{sec3-lag-tor-eq}
&\kappa&(\mathbf{u}^{\tau}\mathbf{T}^{\alpha}_{\beta\tau}+\mathbf{u}^{\tau}\mathbf{T}^{\theta}_{\theta\tau}\delta^{\alpha}_{\beta}
+\mathbf{u}^{\alpha}\mathbf{T}^{\theta}_{\theta\beta})\\
&=&\lambda\sqrt{-\Phi}\epsilon^{\alpha\tau\rho\sigma}\mathbf{v}_{\rho}\mathbf{T}^{\theta}_{\tau\sigma}\mathbf{g}_{\theta\beta}
+\lambda\epsilon^{\alpha\tau\rho\sigma}\partial_{\tau}(\mathbf{v}_{\rho}\sqrt{-\Phi})\mathbf{g}_{\sigma\beta}.\nonumber
\end{eqnarray}
Eqs.~(\ref{sec3-lag-con-met-s}), (\ref{sec3-lag-tor-s}) and (\ref{sec3-lag-tor-eq}) uniquely determine the torsion tensor as
\begin{eqnarray}
\label{sec3-lag-tor-sol}
\mathbf{T}^{\mu}_{\alpha\beta}&=&\frac{1}{2}\mathbf{g}^{\mu\sigma}\mathbf{u}^{\tau}(\mathbf{v}_{\alpha}
\mathbf{\Gamma}_{\tau,\sigma\beta}
-\mathbf{v}_{\beta}\mathbf{\Gamma}_{\tau,\sigma\alpha})\\
&+&\mathbf{v}_{\alpha}\mathbf{K}^{\mu}_{\beta}-\mathbf{v}_{\beta}\mathbf{K}^{\mu}_{\alpha}\nonumber\\
&+&\frac{\lambda}{2\kappa}\sqrt{-\Phi}\mathbf{g}^{\mu\tau}\mathbf{g}^{\theta\sigma}
(\partial_{\tau}\mathbf{v}_{\theta}-\partial_{\theta}\mathbf{v}_{\tau})\mathbf{u}^{\nu}\epsilon_{\nu\sigma\alpha\beta}\nonumber\\
&+&\frac{\kappa}{4\lambda}\frac{1}{\sqrt{-\Phi}}
(\mathbf{g}^{\mu\rho}\mathbf{g}^{\sigma\theta}
-\mathbf{g}^{\mu\sigma}\mathbf{g}^{\rho\theta})\mathbf{u}^{\tau}\mathbf{\Gamma}_{\tau,\rho\theta}
\mathbf{u}^{\nu}\epsilon_{\nu\sigma\alpha\beta}.\nonumber
\end{eqnarray}
where
\begin{eqnarray}
\label{sec3-lag-tor-k}
\mathbf{K}^{\alpha}_{\beta}&=&
\frac{\lambda}{\kappa}
\epsilon^{\alpha\theta\sigma\tau}\mathbf{v}_{\theta}\mathbf{g}_{\sigma\beta}\partial_{\tau}\sqrt{-\Phi}\\
&+&\frac{\lambda}{\kappa}\sqrt{-\Phi}\mathbf{n}^{\alpha}_{\rho}\mathbf{g}_{\sigma\beta}\epsilon^{\rho\sigma\tau\theta}
\partial_{\tau}\mathbf{v}_{\theta}\nonumber\\
&+&\frac{\lambda^2}{\kappa^2}\Phi\mathbf{g}^{\alpha\tau}\mathbf{n}^{\theta}_{\beta}
(\partial_{\tau}\mathbf{v}_{\theta}-\partial_{\theta}\mathbf{v}_{\tau}).\nonumber
\end{eqnarray}
From Eq.~(\ref{sec3-lag-con-re}), we can express the field strength in terms of the curvature
\begin{eqnarray}
\label{sec3-lag-cur}
\mathbf{F}^{k}_{\mu\nu}=\frac{1}{2}\epsilon^{kmn}\mathbf{R}^{\rho}_{\hspace{1mm}\sigma\mu\nu}
\mathbf{e}^{n}_{\rho}\mathbf{E}^{\sigma}_{m},
\end{eqnarray}
and
\begin{eqnarray}
\label{sec3-lag-cur-rie}
\mathbf{R}^{\rho}_{\hspace{1mm}\sigma\mu\nu}=\partial_{\mu}\mathbf{\Gamma}^{\rho}_{\sigma\nu}
-\partial_{\nu}\mathbf{\Gamma}^{\rho}_{\sigma\mu}+\mathbf{\Gamma}^{\rho}_{\mu\tau}\mathbf{\Gamma}^{\tau}_{\sigma\nu}
-\mathbf{\Gamma}^{\rho}_{\nu\tau}\mathbf{\Gamma}^{\tau}_{\sigma\mu}
\end{eqnarray}
is the definition of Riemann tensor. The Lagrangian~(\ref{sec2-lag-c}) can be rewritten as
\begin{eqnarray}
\label{sec3-lag}
\mathscr{L}=-\frac{\kappa}{4}\epsilon^{\alpha\beta\mu\nu}\mathbf{R}_{\alpha\beta\mu\nu}
-\lambda\sqrt{-\Phi}\mathcal{R}+\mathrm{c.c},
\end{eqnarray}
where
\begin{eqnarray}
\label{sec3-lag-rie-sca}
\mathcal{R}=\mathbf{g}^{\alpha\beta}\mathbf{R}^{\tau}_{\hspace{1mm}\alpha\beta\theta}\mathbf{n}^{\theta}_{\tau}
\end{eqnarray}
is a scalar curvature, and
\begin{eqnarray}
\label{sec3-lag-rie-cov}
\mathbf{R}_{\alpha\beta\mu\nu}=\mathbf{g}_{\alpha\tau}\mathbf{R}^{\tau}_{\hspace{1mm}\beta\mu\nu}
\end{eqnarray}
is the covariant Riemann tensor. The second term of Eq.~(\ref{sec3-lag}) is similar to the Einstein-Hilbert Lagrangian. The first term of Eq.~(\ref{sec3-lag}) vanishes in Einstein's gravity, due to the first Bianchi identity, although it contributes in our case because the connection~(\ref{sec3-lag-con-sol}) has the torsion piece~(\ref{sec3-lag-tor-sol}). We define the contracted tensor
\begin{eqnarray}
\label{sec3-lag-ric}
\mathbf{R}^{\alpha}_{\beta}=\mathbf{g}^{\tau\theta}\mathbf{R}^{\alpha}_{\hspace{1mm}\tau\theta\beta}.
\end{eqnarray}

Then the equations of motion in Eq.~(\ref{sec2-lag-eq}) can be rewritten as
\begin{eqnarray}
\label{sec3-lag-eq}
2\lambda\sqrt{-\Phi}(\mathbf{n}^{\alpha}_{\sigma}\mathbf{R}^{\sigma}_{\beta}-\frac{1}{2}\mathcal{R}\delta^{\alpha}_{\beta})
&-&\frac{\kappa}{2}\epsilon^{\alpha\sigma\mu\nu}\mathbf{R}_{\beta\sigma\mu\nu}\nonumber\\
&=&(\mathcal{S}^{\alpha}_{\beta}-\mathrm{c.c}),
\end{eqnarray}
where
\begin{eqnarray}
\label{sec3-lag-eq-def}
\mathcal{S}^{\alpha}_{\beta}&=&\lambda\sqrt{-\Phi}\mathcal{R}\mathbf{u}^{\alpha}\mathbf{v}_{\beta}
-2\lambda\sqrt{-\Phi}\mathbf{u}^{\tau}\mathbf{R}^{\rho}_{\tau}\mathbf{n}^{\alpha}_{\rho}\mathbf{v}_{\beta}\nonumber\\
&-&\frac{\lambda}{2}\frac{1}{\sqrt{-\Phi}}\mathbf{u}^{\tau}\mathbf{R}^{\sigma}_{\tau}
\mathbf{n}^{\rho}_{\sigma}\mathbf{\bar{g}}_{\rho\beta}\mathbf{u}^{\alpha}.
\end{eqnarray}
The left hand of Eq.~(\ref{sec3-lag-eq}) is similar to the Einstein's gravitational equation, although it is complex. Note that only the imaginary part of $\mathcal{S}^{\alpha}_{\beta}$ contributes to the equations of motion.

\section{Classical Solutions}\label{sec4}

We consider classical solutions of Eq.~(\ref{sec3-lag-eq}). For stationary solutions with spherical symmetry, we use the following ansatz for $g_{\mu\nu}$
\begin{eqnarray}
\label{sec4-g}
g_{\mu\nu}dx^{\mu}dx^{\nu}=-p^2(r)dt^2+q^4(r)\bigl(dr^2+r^2d\Omega^2\bigr)
\end{eqnarray}
and for $f_{\mu\nu}$
\begin{eqnarray}
\label{sec4-f}
f_{\mu\nu}dx^{\mu}dx^{\nu}=-2p(r)q^2(r)dtdr+{\beta}\hspace{1mm}q^4(r)r^2d\Omega^2.~~
\end{eqnarray}
where $d\Omega^2=d\theta^2+\sin^2\theta{d}\phi^2$, and $\beta$ is a real constant. The ansatz in Eqs.~(\ref{sec4-g}) and~(\ref{sec4-f}) satisfy the constraint~(\ref{sec1-gf}). Eq.~(\ref{sec3-lag-eq}) is a complex equation, whose real part yields two independent equations
\begin{eqnarray}
\label{sec4-eq-1}
\mathrm{Re}(\mathbf{a})q(r)-8\mathrm{Re}(\mathbf{b})r\frac{dq}{dr}&=&4\mathrm{Re}(\mathbf{b})r^2\frac{d^2q}{dr^2},\\
\label{sec4-eq-2}
\frac{1}{Q}\frac{dQ}{dr}-4\frac{1}{q}\frac{dq}{dr}&=&\frac{1}{p}\frac{dp}{dr},
\end{eqnarray}
where $Q(r)=q^4(r)+2rq^3(r)\frac{dq}{dr}$, and
\begin{eqnarray}
\label{sec4-eq-1-a}
\mathbf{b}&=&2(1+i\beta)(\kappa^2-\lambda^2)\bar{\lambda},\\
\label{sec4-eq-2-b}
\mathbf{a}&=&-\mathbf{b}-8\mathbf{\bar{\lambda}}\lambda^2.
\end{eqnarray}
The solutions for $q(r)$ and $p(r)$ are
\begin{eqnarray}
\label{sec4-eq-1-sol}
q(r)&=&c_{1}r^{\alpha-\frac{1}{2}}+c_{2}r^{-\alpha-\frac{1}{2}},\\
\label{sec4-eq-2-sol}
p(r)&=&-2\alpha{c_3}\frac{c_{2}-c_{1}r^{2\alpha}}{c_{2}+c_{1}r^{2\alpha}},
\end{eqnarray}
where $\alpha=\frac{1}{2\mathrm{Re}(\mathbf{b})}\sqrt{\mathrm{Re}(\mathbf{b})\mathrm{Re}(\mathbf{a+b})}$, and $c_1$, $c_2$ and $c_3$ are three integral constants. We can check that the above solutions also solve the imaginary part of Eq.~(\ref{sec3-lag-eq}), hence $p(r)$ and $q(r)$ in Eqs.~(\ref{sec4-eq-1-sol}) and~(\ref{sec4-eq-2-sol}) are solutions of Eq.~(\ref{sec3-lag-eq}). When $\alpha$ is real, $c_1$ and $c_2$ are required to be real;  in this case, $c_1$ and $c_3$ can be absorbed into the redefinition of $r$ and $t$. When $\alpha$ is imaginary, $c_1$ and $c_2$ are required to be conjugate complex numbers;  in this case, the real~(or imaginary) part of $c_1$ and $c_3$ can be absorbed into the redefinition of $r$ and $t$. In both cases, only one effective parameter is left. With $\alpha$, the metric in Eq.~(\ref{sec4-eq-1-sol}) has two hairs. An interesting case happens when
\begin{eqnarray}
\label{sec4-eq-1-sol-ss-c}
\alpha=\frac{1}{2},~~c_1=1,~~c_2=\frac{1}{4}r_{\mathrm{s}},~~c_3=1.
\end{eqnarray}
We obtain, from Eq.~(\ref{sec4-g})
\begin{eqnarray}
\label{sec4-g-ss}
&-&\left(\frac{1-\frac{r_{\mathrm{s}}}{4r}}{1+\frac{r_{\mathrm{s}}}{4r}}\right)^2dt^2
+\biggl(1+\frac{r_{\mathrm{s}}}{4r}\biggr)^4\bigl(dr^2+r^2d\Omega^2\bigr)\nonumber\\
&=&g_{\mu\nu}dx^{\mu}dx^{\nu}.
\end{eqnarray}
This is the Schwarzschild metric in isotropic coordinates, where $r_{\mathrm{s}}$ is the Schwarzschild radius. For $\alpha=\frac{1}{2}$, we need $\mathrm{Re}(\mathbf{a})=0$, which determines $\beta$
\begin{eqnarray}
\label{sec4-g-ss-beta}
\beta=\frac{\mathrm{Re}\bigl(\kappa^2\bar{\lambda}+3\lambda^2\bar{\lambda}\bigr)}
{\mathrm{Im}\bigl(\kappa^2\bar{\lambda}-\lambda^2\bar{\lambda}\bigr)}.
\end{eqnarray}

Now we consider time-dependent solutions. The ansatz for $g_{\mu\nu}$ is
\begin{eqnarray}
\label{sec4-g-t}
g_{\mu\nu}dx^{\mu}dx^{\nu}=-dt^2+a^2(t)dx_idx_i,
\end{eqnarray}
and $f_{\mu\nu}$ is
\begin{eqnarray}
\label{sec4-f-t}
f_{\mu\nu}dx^{\mu}dx^{\nu}&=&f_{00}dt^2+2f_{0i}dtdx_i+f_{ij}dx_idx_j,\\
f_{\mu\nu}&=&\begin{pmatrix}
f_{00}&f_{01}&f_{01}&f_{01}\\
f_{01}&f_{11}&f_{12}&f_{12}\\
f_{01}&f_{12}&f_{11}&f_{12}\\
f_{01}&f_{12}&f_{12}&f_{11}\\
\end{pmatrix},\nonumber
\end{eqnarray}
in which
\begin{eqnarray}
\label{sec4-f-t-com}
f_{00}&=&\frac{1}{2}\bigl(\omega-\frac{1}{\omega}\bigr),~
f_{11}=\frac{1}{2}\bigl(\frac{\omega}{3}-\frac{1}{\omega}\bigr)a^2(t),\\
f_{12}&=&\frac{\omega}{6}a^2(t),~~f_{01}=\frac{1}{2\sqrt{3}}\bigl(\omega+\frac{1}{\omega}\bigr)a(t).\nonumber
\end{eqnarray}
Here $\omega$ is a real constant. The left hand of Eq.~(\ref{sec3-lag-eq}) yields two independent expressions
\begin{eqnarray}
\label{sec4-t-eq-1}
-\frac{1+2i\omega}{8\lambda\omega^2}\bigl((-i+\omega)^2\kappa^2+(i+\omega)^2\lambda^2\bigr)a(\dot{a}^2+2a\ddot{a}),~~\\
\label{sec4-t-eq-2}
\frac{1}{4\sqrt{3}\lambda\omega^2}(1+\omega^2)(1+2i\omega)\bigl(\kappa^2+\lambda^2\bigr)(\dot{a}^2-a\ddot{a}),~~
\end{eqnarray}
where $\dot{a}=\frac{da}{dt}$. There are $4$ independent expressions for the right hand of Eq.~(\ref{sec3-lag-eq}). Eq.~(\ref{sec3-lag-eq}) can be satisfied if
\begin{eqnarray}
\label{sec4-t-eq-a}
(-i+\omega)^2\kappa^2+(i+\omega)^2\lambda^2&=&0,\\
\label{sec4-t-eq-b}
\dot{a}^2-a\ddot{a}&=&0.
\end{eqnarray}
The solution of Eq.~(\ref{sec4-t-eq-b}) is
\begin{eqnarray}
\label{sec4-t-eq-b-sol}
a(t)=e^{Ht},
\end{eqnarray}
where $H$ is a integral constant. Eq.~(\ref{sec4-t-eq-a}) is a complex equation, whose solution yields
\begin{eqnarray}
\label{sec4-t-eq-a-sol-1}
~~&~~&\hspace{12mm}\lambda\bar{\lambda}=\kappa\bar{\kappa},\\
\label{sec4-t-eq-a-sol-2}
\omega&=&\frac{\kappa\bar{\lambda}+\lambda\bar{\kappa}}{\vert{\kappa+i\lambda}\vert^2},~\mathrm{or}~
\omega=-\frac{\kappa\bar{\lambda}+\lambda\bar{\kappa}}{\vert{\kappa-i\lambda}\vert^2},
\end{eqnarray}
where $\vert{z}\vert$ is the module of the complex number $z$. For this solution, $g_{\mu\nu}$ is the de Sitter metric. For the existence of this solution, the couplings $\lambda$ and $\kappa$ are required to satisfy the constraint~(\ref{sec4-t-eq-a-sol-1}).

Now we consider the solution of constant background. Obviously, any constant value of $g_{\mu\nu}$ and $f_{\mu\nu}$ which satisfy the constraint~(\ref{sec1-gf}) are solutions of Eq.~(\ref{sec3-lag-eq}). For the solution in Eqs.~(\ref{sec4-eq-1-sol}) and (\ref{sec4-eq-2-sol}), in the case that the parameters are given by Eq.~(\ref{sec4-eq-1-sol-ss-c}), $p(r)$ and $q(r)$ are constant when $r\rightarrow\infty$. We see that $g_{\mu\nu}$ is the Lorentz metric in this limit, but $f_{\mu\nu}$ breaks Lorentz invariance. For the solution in Eqs.~(\ref{sec4-g-t}) and (\ref{sec4-f-t}), when $a(t)=1$, the metrics are constant, and they solve Eq.~(\ref{sec3-lag-eq}). We see that $g_{\mu\nu}$ is the Lorentz metric, but $f_{\mu\nu}$ breaks Lorentz invariance. Hence the constant background breaks Lorentz invariance spontaneously.

\section{Discussions}\label{sec5}

In this paper, we have provided a gravitational model in terms of the adjoint frame field $e^{IJ}_{\mu}$. This model describes interactions between two metrics. In section~\ref{sec2}, using the $SO(3,\mathbb{C})$ variables, we construct a concise Lagrangian with $2$ complex coupling constants. In section~\ref{sec3}, we give the metric-like formulations of the Lagrangian and equations of motion. We also obtain the Schwarzschild solution and the de Sitter solution in section~\ref{sec4}.

The black hole solution in Eq.~(\ref{sec4-eq-1-sol}) has two effective hairs, which reduces to the Schwarzschild solution in a special case. The stability, uniqueness, and thermodynamical properties of this black hole solution are of theoretical interest. The $\alpha$ hair of this solution shall correct the geodesic equations, and its value can be restricted by the experimental data from the perihelion precession of Mercury, the deflection of light by the sun and the gravitational redshift.

The coupling constants $\kappa$ and $\lambda$ is required to satisfy the constraint~(\ref{sec4-t-eq-a-sol-1}) for the existence of de Sitter solution. It remains to be answered wether this constraint depends on our ansatz~(\ref{sec4-g-t}) and (\ref{sec4-f-t}). The existence of de Sitter solution without cosmological constant supports that our bimetric gravity model could have the capability to interpret the cosmological acceleration. To obtain a realistic cosmological model, the matter energy-momentum tensor is required to be plugged into Eq.~(\ref{sec3-lag-eq}). Because Eq.~(\ref{sec3-lag-eq}) is a complex equation, an additional energy-momentum tensor besides the conventional one could be required to ensure the consistency of the equation.

Bimetric gravity generally suffers from a ghost problem~\cite{Boulware:1973my} and the vDVZ discontinuity problem~\cite{vanDam:1970vg,Zakharov:1970cc}. A detailed analysis is required in order to see wether our bimetric gravity model is free from these problems.

\begin{acknowledgments}

This work was supported in part by Fondecyt~(Chile) grant 1140390 and by Project Basal under Contract No.~FB0821.

\end{acknowledgments}

\bibliographystyle{apsrev4-1}
\bibliography{cmRef}

\end{document}